\documentclass{article}
\usepackage{graphicx}
\graphicspath{ {./} }

\PassOptionsToPackage{numbers}{natbib}

\usepackage[preprint]{neurips_2022}
\usepackage{makecell, multirow}

\usepackage[utf8]{inputenc} 
\usepackage[T1]{fontenc}    
\usepackage{hyperref}       
\usepackage{url}            
\usepackage{booktabs}       
\usepackage{amsfonts}       
\usepackage{nicefrac}       
\usepackage{microtype}      
\usepackage{xcolor}         

\title{MolE: a molecular foundation model for drug discovery}

\author{%
  Oscar M\'endez-Lucio\\
  Recursion\\
  \texttt{oscar.mendez-lucio@recursion.com} \\
  \And
  Christos Nicolaou\\
  Recursion\\
  \texttt{c.nicolaou@recursion.com} \\
  \And
  Berton Earnshaw\\
  Recursion\\
  \texttt{berton.earnshaw@recursion.com} \\
}

\begin{document}

\maketitle

\begin{abstract}
  Models that accurately predict properties based on chemical structure are valuable tools in drug discovery. However, for many properties, public and private training sets are typically small, and it is difficult for the models to generalize well outside of the training data. Recently, large language models have addressed this problem by using self-supervised pretraining on large unlabeled datasets, followed by fine-tuning on smaller, labeled datasets \cite{Brown2020-xw, Zhang2022-ct, Chowdhery2022-zr}. In this paper, we report MolE, a molecular foundation model that adapts the DeBERTa \cite{He2020-uy} architecture to be used on molecular graphs together with a two-step pretraining strategy. The first step of pretraining is a self-supervised approach focused on learning chemical structures, and the second step is a massive multi-task approach to learn biological information. We show that fine-tuning pretrained MolE achieves state-of-the-art results on 9 of the 22 ADMET tasks included in the Therapeutic Data Commons \cite{Huang2021-vj}.
\end{abstract}

\section{Introduction}

Machine learning has been successfully applied to chemical sciences for many decades \cite{Martin2012-ru}. In particular, molecular property prediction has been critical in successfully advancing material and drug discovery projects \cite{Butler2018-qf}. Nonetheless, a major challenge in this area is to represent a molecule in a way that is compatible with machine learning algorithms with minimum information loss. Initially, molecules were represented in terms of their physicochemical properties (e.g., partition coefficient) or information that can be obtained from the molecular formula such as molecular weight or number of heteroatoms \cite{Willett1998-wp}. While this approach was successful for the first quantitative structure-activity relationship (QSAR) studies \cite{Hansch1964-oj}, it uses global properties of the molecule that do not provide information about the topology of the molecular graph. With time, molecules were described in more sophisticated ways using molecular fingerprints such as MACCS keys \cite{Durant2002-wo} and Extended Connectivity Fingerprints (ECFPs) \cite{Rogers2010-mg}. These molecular fingerprints encode substructures of the molecules either in the form of preset chemical groups or as atom environments.

Following recent advances in natural language modeling, it was noted that molecules could be used directly as input for predictive models in the form of SMILES \cite{Weininger1988-yv, Weininger1989-ro}, a string-based representation developed to store and search molecular structures in a fast and easy way. SMILES have been used as inputs for deep learning architectures such as recurrent neural networks (RNNs) \cite{Winter2019-zy} and Transformers \cite{Vaswani2017-ke}, though they suffer from the fact that molecules do not have unique SMILES representations. Other types of string-based representations have been proposed \cite{OBoyle2018-mc, Krenn2020-hi}, e.g., Self-Referencing Embedded Strings (SELFIES), which encodes the molecular graph in the form of a Chomsky type-2 free context grammar appropriate for deep learning applications.

An alternative is to use graph representations of the molecule where nodes represent atoms and edges represent bonds. Such an approach is compatible with graph neural networks (GNNs), which have been extensively used for molecular property prediction \cite{Duvenaud2015-zf, Yang2019-bl}. Typically, GNNs aggregate the local information of each node with that of its neighboring atoms, and this information is then aggregated into a single molecular representation used to predict specific properties. Despite the fact that GNNs could provide the most natural way for learning representations of molecules that perform well in property prediction tasks, they suffer from some drawbacks, namely that each atom can aggregate information only from nearest neighbors.

In this paper, we show how architectures developed for natural language processing can be used with molecular graph representations for property prediction. In particular, we present \emph{MolE}, a model that learns molecular embeddings at the atomic environment level, directly from molecular graphs. MolE is based on DeBERTa \cite{He2020-uy}, a transformer architecture that uses disentangled attention in order to account for relative token positions. We also describe a pretraining strategy that combines self-supervised pretraining on chemical structures with supervised multi-task pretraining of pharmacological and other biological properties, yielding a molecular foundation model capable of being fine-tuned on small datasets to achieve top performance on typical benchmark tasks. This work is of relevance for chemical sciences where large amounts of molecular structures are available but the size of labeled datasets is usually very small.

\section{Related work}
\subsection{Molecular fingerprints}

In the last decades, the most common way to encode molecules for machine learning models is using molecular fingerprints. Among the most popular are MACCs keys \cite{Durant2002-wo} and ECFPs \cite{Rogers2010-mg}. MACCs keys encode a molecule into a binary vector usually of 166 bits though they can be longer if needed. Each bit encodes the presence or absence of a particular pre-defined chemical group or substructure in the molecule. In a similar way, ECFPs encode molecular substructures into a binary vector of fixed length using the Morgan algorithm \cite{Morgan1965-ym}. An alternative to ECFPs are functional-class fingerprints (FCFPs) \cite{Rogers2010-mg} which use pharmacophoric information for atom identifiers, namely, hydrogen-bond acceptor/donor, negatively/positively ionizable, aromatic, and halogen. This is a more abstract form of fingerprint which represents equivalent chemical groups (e.g., halogens) in the same way irrespective of the atom type resulting in functional atom environments.

Despite the fact that ECFPs and FCFPs represent the molecular structure in an appropriate way for similarity search and have been successfully used for machine learning, they present some issues. The use of a hash function to project them into a defined length bit vector causes ‘collisions’, meaning that two atom environments can be mapped to the same bit. The number of collisions and hence information loss increases as the length of the bit vector decreases \cite{MacCuish2001-gh}.

\subsection{Molecular property prediction}

Different machine learning methods have been used for molecular property prediction. For example, models like multilayer perceptrons and XGBoost, in combination with molecular fingerprints, have been used to successfully predict bioactivity and toxicity of molecules \cite{Mayr2016-jb, Mayr2018-we, Tian2022-fy}. However, more suitable architectures such as message passing neural networks (MPNNs), a form of graph neural network (GNN), have shown improvements in different tasks such as the prediction of biological activity, toxicity, and quantum and physicochemical properties \cite{Yang2019-bl}. Also, several approaches have taken advantage of NLP architectures (e.g., RNNs and Transformers) since molecules can be represented by SMILES \cite{Winter2019-zy, Fabian2020-zk, Chithrananda_undated-eu, Ahmad2022-qx, Wang2019-kq}.

Despite these models being trained and tested on similar datasets, it is difficult to compare them in a robust and fair manner that represents their performance in real-world applications \cite{Huang2021-vj}. To address this difficulty, frameworks like MoleculeNet \cite{Wu2018-tc} and DeepChem \cite{Ramsundar2019-wy} compiled various datasets and tasks for use as molecular benchmarks. More recently, Huang et al. developed the Therapeutic Data Commons (TDC) \cite{Huang2021-vj} a framework for systematic evaluation of machine learning models across a variety of pharmacological tasks. All models reported in this paper were benchmarked against the ADMET tasks and datasets in TDC.

\subsection{Self-supervised pretraining for chemistry models}

Though pretraining a model on large unlabeled datasets is a common and successful practice for language modeling, it is less common for molecular models. Some approaches \cite{Fabian2020-zk, Chithrananda_undated-eu, Ahmad2022-qx, Wang2019-kq} use a masked language model similar to BERT \cite{Devlin2018-no} for predicting the identity of masked tokens of a SMILES string based on the rest of the SMILES context. After pretraining, the model is fine-tuned for different downstream tasks. A different approach based on SMILES was presented by Winter et al. \cite{Winter2019-zy} where they trained an autoencoder on 72 million molecules and used the latent space embeddings as inputs for a multilayer perceptron.

On the other hand, pretraining strategies for molecular graphs are not as straightforward. For example, Hu et al. pretrained GNNs using context prediction or attribute masking \cite{Hu2019-jd}. Context prediction uses a binary classifier to indicate whether a particular atom environment corresponds to a particular context graph (i.e., nodes beyond the atom environment). In attribute masking, random nodes are masked and the task is to predict their attributes such as atom type or chirality. Wang et al. proposed an alternative pretraining strategy based on contrastive learning \cite{Wang2022-yy}. They applied different augmentation strategies (e.g., atom masking, bond deletion, subgraph removal) and generated two correlated molecular graphs for each molecule in the training set. They proceeded to encode each augmented graph with a GNN using a contrastive loss to maximize the agreement between the two latent vectors corresponding to the same parent molecule. This approach seems suitable for learning meaningful embeddings for the whole molecule and not just at the  atomic level.

\section{Method}
\subsection{Model}

Transformer architectures are powerful models for performing NLP tasks \cite{Vaswani2017-ke}. However, they need to be adapted for tasks that are invariant to input order. A solution to this problem is to remove the typical positional encoding used and instead use relative position embeddings. For this reason, we chose to use a variation of the DeBERTa model \cite{He2020-uy}. DeBERTa is a transformer architecture in which attention weights include information about a token's content and its relative position to other tokens. In fact, a standard self-attention layer consists of queries, keys and values $Q,K,V\in R^{N \times d}$, and is calculated as

\begin{equation}
\begin{array}{l}
    H_0= \mathrm{softmax}(A)V, \quad A=\frac{QK^T}{\sqrt{d}}
\end{array}
\end{equation}

where $H_0 \in R^{N \times d}$ is the output hidden vectors after self-attention, and $d$ the hidden dimension. In DeBERTa, the disentangled self attention incorporates relative position information for each $K$ and $Q$ pair:

\begin{equation}
\begin{array}{l}
a_{ij} = Q_i^c {K_j^c}^T + Q_i^c {K_{i,j}^p}^T + K_j^c {Q_{j,i}^p}^T \\
\\
H_0= \mathrm{softmax}(A)V^c, \quad A=\frac{a}{\sqrt{3d}}
\end{array}
\end{equation}

where $Q^c,K^c,V^c\in R^{N \times d}$ are context queries, keys and values containing information about the token (similar to standard self-attention) and $Q_{i,j}^p,K_{i,j}^p\in R^{N \times d}$ to encode the relative positions of tokens $i$ and $j$.

\subsection{Pretraining strategy}

\begin{figure*}[ht]
\vskip 0.2in
\begin{center}
\centerline{\includegraphics[width=\textwidth]{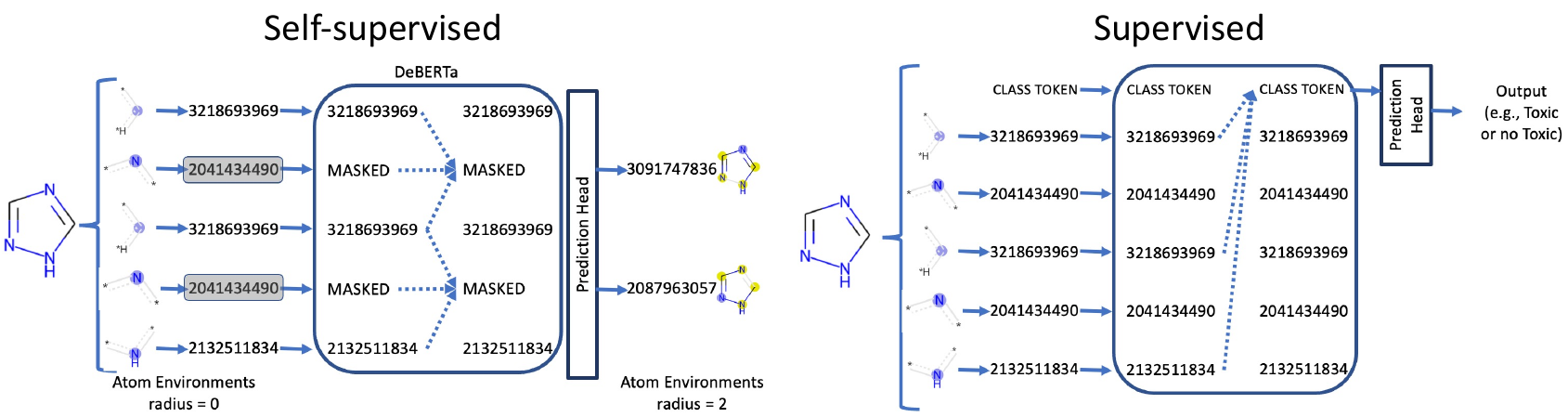}}
\caption{Pretraining methods. a) A self-supervised approach in which an input atom is masked and the task is to predict its radius-2 atom environment. Note that in this particular example the two masked tokens are the same (2041434490), but the atom environment associated with each is different. b) A supervised approach in which a class token is added to the input and predictions are made using its final embeddings.}
\label{fig1}
\end{center}
\vskip -0.2in
\end{figure*}

As mentioned, when combined with fine-tuning, self-supervised pretraining is a good alternative to transfer information from large unlabeled datasets to smaller datasets with labels. Here we present a two-step pretraining strategy (see Figure \ref{fig1}). The first step is a self-supervised approach to learn to represent chemical structures. For this we use a BERT-like procedure in which each atom is randomly masked with a probability of 15\%, from which 80\% of the selected tokens are replaced by a mask token, 10\% replaced by a random token from the vocabulary, and 10\% are not changed. Different from BERT, the prediction task is not to predict the identity of the masked token, but to predict the corresponding atom environment (or functional atom environment) of radius 2, that is, all atoms and their associated bonds that are separated from the masked atom by two or less bonds. The motivation of this step is to incentivize the model to aggregate information from neighboring atoms when learning embeddings of local molecular features. The second step uses a graph-level pretraining in a supervised way with a large labeled dataset. As proposed by Hu et al. \cite{Hu2019-jd}, combining node- and graph-level pretraining helps to learn local and global features that improve the final prediction performance.

\subsection{Molecular representation}

The way molecules are represented is of great importance for property prediction. The fact that we use a language model does not mean we are required to use a string-based representation of molecules such as SMILES \cite{Weininger1988-yv, Weininger1989-ro} or SELFIES  \cite{Krenn2020-hi}. In fact, the way in which  DeBERTa calculates disentangled attention allows MolE to directly work with graphs by providing node feature information at the token level and graph connectivity via the relative position information. In this study, we used the Morgan algorithm as implemented in RDKit \cite{Landrum_undated-xd, Morgan1965-ym} to calculate atom environments. Each atom in the molecule is initially represented by the identifier of its atom environment of radius 0. This contains information regarding the atom and all bonds attached to it, without including information of neighboring atoms. We used the same strategy to generate atom environments and functional atom environments of radius 2 for use as labels.

In addition to tokens, DeBERTa also takes relative position information as input which is an important inductive bias since it gives information about connectivity. In the case of MolE, the graph connectivity is given as a distance matrix where the $i,j$th entry corresponds to the length of the shortest path of bonds between atoms $i$ and $j$.

\section{Experiments}
\subsection{Datasets}

The self-supervised pretraining was done using the GuacaMol \cite{Brown2018-ke} dataset, which was previously used to train MolBERT \cite{Fabian2020-zk} and consists of $\sim$1.2 million molecules for training and a validation set of $\sim$79K molecules, mostly extracted from ChEMBL \cite{Gaulton2012-hc}. It is worth mentioning that only molecules with no more than 100 heavy atoms were used, and we removed from the training set all molecules included in TDC test sets to avoid information leakage. All remaining SMILES were transformed into molecular graphs using RDKit from which distance matrices and atom environments were calculated. Atom environment identifiers were aggregated into two vocabularies, one used for input and one for labels. The input vocabulary consists of 207 tokens corresponding to all atom environments of radius 0 present in the 1.2 million molecules in GuacaMol, plus the $\sim$880 million molecules in ZINC20 \cite{Irwin2020-db}. Similarly the vocabulary used for labels contains $\sim$141K atom environments or $\sim$114K functional atom environments (Table \ref{tab1}). These were selected taking the 90K most frequent atom environments or functional environments from GuacaMol training set plus the 90K most frequent from ZINC20 and removing those that appear in less than 3 molecules.

The supervised pretraining was done using $\sim$456K molecules with activity data on 1,310 prediction tasks from ChEMBL, which was curated following the protocol proposed by Mayr et al. \cite{Mayr2018-we} and used for pretraining by Hu et al. \cite{Hu2019-jd}. Here again we removed $\sim$9,900 molecules that were present in the test sets of the benchmark datasets.

\begin{table}[!htp] 
    \caption{Summary of tasks and input/output vocabularies used for self-supervised and supervised pretraining.} 
    \label{tab1}
    \centering
    \begin{tabular}{lccccc}
        \toprule
        &\multicolumn{2}{c}{Self-supervised pretraining} & &\multicolumn{2}{c}{Supervised pretraining} \\
        \cmidrule{2-3}\cmidrule{5-6}
        Label type &Input (Radius 0 ) &Output (Radius 2 ) & &Molecules &Tasks \\
        \midrule
        Atom Envs &207 &~141K & &~456K &1,310 \\
        Functional Envs &207 &~114K & &~456K &1,310 \\
        \bottomrule
    \end{tabular}
\end{table}

\subsection{Training}

\begin{figure*}[ht]
\vskip 0.2in
\begin{center}
\centerline{\includegraphics[width=\textwidth]{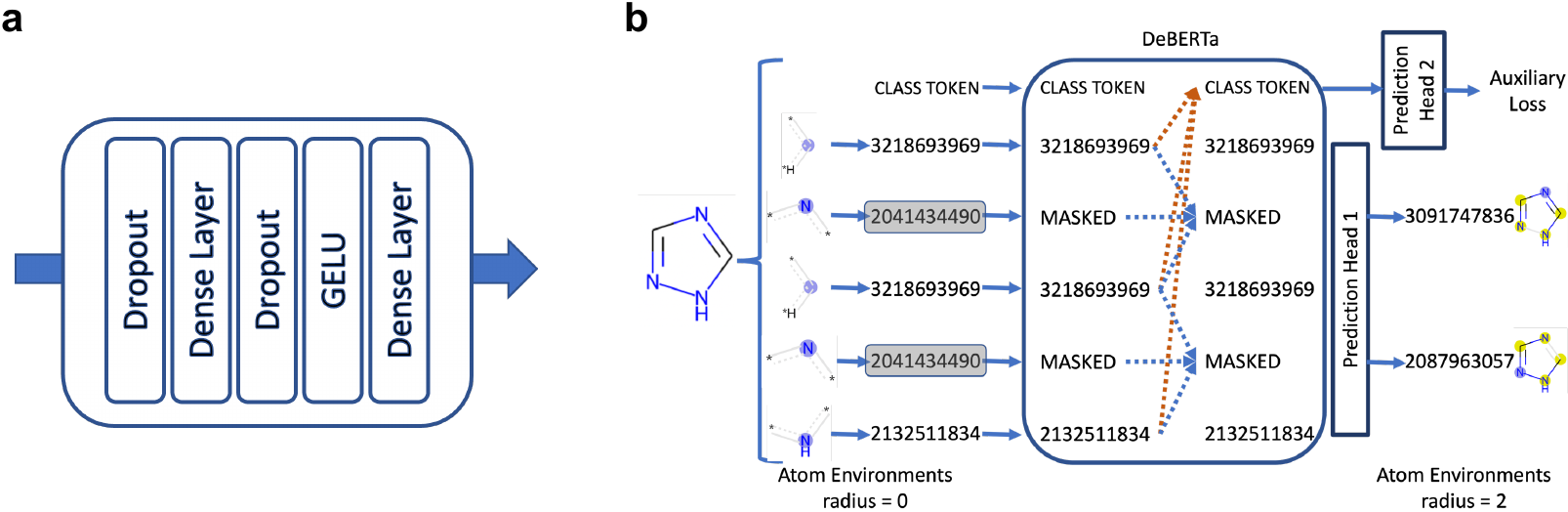}}
\caption{a) Representation of the prediction head used in this study. b) Representation of the self-supervised approach  trained with an auxiliary loss. We evaluated two different auxiliary losses: logP prediction and fingerprint prediction.}
\label{fig2}
\end{center}
\vskip -0.2in
\end{figure*}

MolE uses the DeBERTa base configuration (12 transformer layers with 12 attention heads each) with a prediction head connected to the output of a class token composed of a two-layer MLP with a GELU and dropout in between (Figure \ref{fig2}a). Models were pretrained for 60,000 steps using a batch size of 512 molecules in both the supervised and self-supervised cases. In case of the self-supervised models, the learning rate was increased to $2\times 10^{-4}$ during the first 10,000 warm up steps, followed by a linear decaying learning rate schedule. For the supervised pretraining, we used a learning rate of $5\times 10^{-6}$ with the same schedule as the self-supervised training. Gradient norms were clipped at 1.0 and no weight decay was used. 

For fine-tuning, only the weights of the prediction head were randomly initialized. Models were trained for 100 epochs using a batch size of 32 molecules. We ran hyperparameter optimization to find the best learning rate (1e-5, 8e-6, 5e-6, 3e-6, 1e-6, 5e-7) and dropout rate (0, 0.1, 0.15 in the prediction head) with a 5-fold cross validation using the folds provided in the TDC benchmark datasets selected via scaffold splitting. During training, the learning rate was linearly increased during the first 10\% of the training steps, and then was kept constant.

\subsection{Benchmark}

MolE models were evaluated using the ADMET benchmark group from the Therapeutic Data Commons (TDC) \cite{Huang2021-vj}. This benchmark provides datasets that have been previously standardized and divided into training and test sets (80\%/20\% using scaffold splitting) to fairly evaluate molecular property prediction models. It is composed of 22 different classification and regression tasks for properties relevant to drug discovery. For example, cell permeability (Caco-2), Human Intestinal Absorption (HIA), and p-glycoprotein inhibition (Pgp), together with other physicochemical properties, can give a good estimate of how much of the drug will be absorbed by the body. Properties like volume of distribution (VDss), the plasma protein binding rate (PPBR), and the Blood-Brain Barrier (BBB), give us an idea of how the drug will be distributed across the body. Knowing whether a molecule inhibits or is substrate for a particular cytochrome (CYP) isoform indicates possible biotransformations that can affect the time the drug remains in the body, which is measured by half-life and clearance rate. Finally, knowing whether a molecule is cardiotoxic (hERG), genotoxic (Ames), or hepatotoxic (DILI) is of great importance in assessing the safety of compounds in humans. More information about each of these tasks is listed in Table \ref{tab2}.

\begin{table}[!htp] 
    \caption{Description of the TDC ADMET datasets and tasks.} 
    \label{tab2}
    \centering
    \begin{tabular}{llcr}
        \toprule
        &Dataset &Metric &Size \\
        \midrule
        \multirow{6}{*}{Absorption} &Caco2 &MAE &906 \\
        &HIA &AUROC &578 \\
        &Pgp &AUROC &1,212 \\
        &Bioavailability &AUROC &640 \\
        &Lipophilicity &MAE &4,200 \\
        &Solubility &MAE &9,982 \\
        \cmidrule(r){1-4}
        \multirow{3}{*}{Distribution} &BBB &AUROC &1,975 \\
        &PPBR &MAE &1,797 \\
        &VDss &Spearman &1,130 \\
        \cmidrule(r){1-4}
        \multirow{6}{*}{Metabolism} &CYP2D6 inhibition &AUPRC &13,130 \\
        &CYP3A4 inhibition &AUPRC &12,328 \\
        &CYP2C9 inhibition &AUPRC &12,092 \\
        &CYP2D6 substrate &AUPRC &664 \\
        &CYP3A4 substrate &AUROC &667 \\
        &CYP2C9 substrate &AUPRC &666 \\
        \cmidrule(r){1-4}
        \multirow{3}{*}{Excretion} &Half life &Spearman &667 \\
        &Clearance microsome &Spearman &1,102 \\
        &Clearance hepatocyte &Spearman &1,020 \\
        \cmidrule(r){1-4}
        \multirow{4}{*}{Toxicity} &hERG &AUROC &648 \\
        &Ames &AUROC &7,255 \\
        &DILI &AUROC &475 \\
        &LD50 &MAE &7,385 \\
        \bottomrule
    \end{tabular}
\end{table}

TDC maintains a leaderboard of the performance of different models on these tasks. These models provide a baseline for performance comparison purposes since they use different architectures and encoding strategies, e.g., pre-calculated descriptors such as Morgan or RDKit 2D fingerprints \cite{Landrum_undated-xd}, CNNs trained using SMILES, and different flavors of graph-based approaches such as NeuralFP \cite{Duvenaud2015-zf}, GCNs \cite{Kipf2016-je}, AttentiveFP \cite{Xiong2020-cx}, and others. It also includes models pretrained with different strategies e.g., AttrMasking and ContextPred \cite{Hu2019-jd}.

\section{Results}
\subsection{Effect of pretraining strategies}

Table \ref{tab3} and Appendix \ref{ext_results} shows the benchmark results of all models pretrained with different strategies compared to the best model reported on the TDC leaderboard. We also present the results of MolE without any pretraining, which was used as a baseline. Interestingly, the model without pretraining already exhibits better results on 4 of the 22 tasks compared to models reported by Huang et al. \cite{Huang2021-vj}. This suggests that just the use of transformers with disentangled attention already positively impacts the predictive power of the model despite fine-tuning on small datasets. Nevertheless, these models did not perform better than the best models reported on the leaderboard (as of September 2022) in any of the predictive tasks. The same table also shows results after self-supervised pretraining using two approaches, namely, MolE (AtomEnvs) and MolE (FunctionalEnvs). In the former, the pretraining task is to predict the atom environment of radius 2 taking as input an atom environment of radius 0. The latter is similar, but now the pretraining task is to predict the functional atom environment of radius 2. High accuracy (> 98\%) on a validation set was obtained for the masked token prediction task with either pretraining strategy. However, MolE (AtomEnvs) performed slightly better on the benchmark tasks, where it outperformed previous models \cite{Huang2021-vj, Tian2022-fy} on the leaderboard on 6 tasks, whereas MolE (FunctionalEnvs) did so on only 3. This is not an unexpected result since we expect pretraining using atom environments to provide a better representation of the molecule compared to the less-specific description provided by functional environments.

We noticed that supervised pretraining helps to improve prediction power even when used without any self-supervised pretraining. Nonetheless the improvement is marginal compared to not using any pretraining at all. We hypothesize that it is hard to learn a transferable representation of molecules at the same time as the prediction task, especially with a small number of examples, and for that reason the gain from using supervised pretraining alone is unremarkable. In a similar way, using supervised learning after the self-supervised approach helped MolE (AtomEnvs) outperform previously reported models on 9 of the 22 tasks. A similar improvement in performance was observed after adding multi-task supervision to MolE (FunctionalEnvs). However, we note that using the supervised approach had a negative effect on tasks related to toxicity where performance decreased.

\begin{table}[!htp] 
    \caption{Number of ADMET tasks in which fine-tuned MolE achieves SOTA performance according to the TDC leaderboard (as of September 2022). Complete results can be found in Appendix \ref{ext_results}.}
    \label{tab3}
    \centering
    \scriptsize
    \begin{tabular}{lccccc}
    \toprule
    Pretraining &Self-Supervised label &No Auxiliary Loss &logP as Aux. Loss &FP as Aux. Loss \\
    \midrule
    None &—- &0 &—- &—- \\
    \midrule
    Only Supervised &—- &3 &—- &—- \\
    \midrule
    \multirow{2}{*}{Only Self-Supervised} &AtomEnvs &6 &4 &3 \\
    &FunctionalEnvs &3 &5 &4 \\
    \midrule
    \multirow{2}{*}{Self-Supervised + Supervised} &AtomEnvs &9 &6 &7 \\
    &FunctionalEnvs &7 &7 &7 \\
    \bottomrule
\end{tabular}
\end{table}

\subsection{Using auxiliary losses in pretraining}

We also explored the idea of using auxiliary losses during self-supervised pretraining. For this we explored two different approaches: learning the partition coefficient (logP) and learning a binary fingerprint (FP) of the molecule. In the first task, we take the standard approach of adding a class token to the input and predict logP from its embeddings (Figure \ref{fig2}b). Here, logP values are calculated for every molecule using RDKit \cite{Landrum_undated-xd}. Table \ref{tab3} shows the results of this task for both MolE (AtomlEnvs) and MolE (FunctionalEnvs). In general, only marginal improvements were observed for the self-supervised version of MolE (FunctionalEnvs), which is best on 5 of 22 tasks. Interestingly, using this auxiliary loss decreased the performance of MolE (AtomEnvs) with both pretrainings.

For the second task, we framed the fingerprint learning as a multitask binary classification problem where the task is to identify the presence or absence of each atom environment in the vocabulary. Results are again shown in Table \ref{tab3}, where it can be observed that fine-tuning performance improved only the self-supervised only version of MolE (FunctionalEnvs), outperforming the leaderboard models in 4 of 22 tasks (compared to 3 without using an auxiliary loss). Our hypothesis is that MolE (FunctionalEnvs) takes greater advantage of this approach to overcome the imprecise nature of functional environments. Interestingly, using an auxiliary loss during self-supervision did not improve the models when supervised pretraining is also used.

\section{Conclusion}

In this paper we described MolE, which uses a transformer with disentangled attention (DeBERTa) to predict chemical and biological properties directly from molecular graphs. The specific contributions of this paper are:
\begin{itemize}
\item {Showing that Transformers can use molecular graphs as input when atom environments are the tokens and relative position embeddings are used.}
\item {Proposing a new, powerful self-supervised approach for training molecular graphs where we predict atom environments of radius > 0 from atom environments of radius 0, which only include information about a single atom and all bonds attached to it.}
\item {By using a two-step (self-supervised then supervised) pretraining approach, producing a molecular foundation model that achieves current state-of-the-art performance on 9 of the 22 ADMET tasks included in the Therapeutic Data Commons(Table \ref{tab4}).}
\end{itemize}

\begin{table}[!htp] 
    \caption{Comparison between the best models reported in the TDC leaderboard (as of September 2022) and fine-tuned MolE.} 
    \label{tab4}
    \centering
    \scriptsize
    \begin{tabular}{lrclccccc}
        \toprule
        &\multirow{2}{*}{Dataset} &\multirow{2}{*}{Metric} &\multicolumn{2}{c}{Best in TDC Leaderboard (September 2022)} & &\multicolumn{2}{c}{MolE (AtomEnvs) + Supervised} \\
        \cmidrule{4-5}\cmidrule{7-8}
        & & &Current Best Model &Result & &Result &Rank \\
        \midrule
\multirow{6}{*}{Absorption} &Caco2 &MAE &XGBoost &0.291 ± 0.015 & &0.310 ± 0.010 &2 \\
&HIA &AUROC &XGBoost &0.988 ± 0.002 & &0.963 ± 0.019 &6 \\
&Pgp &AUROC &SimGCN &0.929 ± 0.010 & &0.915 ± 0.005 &5 \\
&Bioavailability &AUROC &SimGCN &0.748 ± 0.033 & &0.654 ± 0.028 &5 \\
&Lipophilicity &MAE &ContextPred &0.535 ± 0.012 & &\underline{\textbf{0.469 ± 0.009}} &1 \\
&Solubility &MAE &XGBoost &0.734 ± 0.006 & &0.792 ± 0.005 &3 \\
\cmidrule{1-8}
\multirow{3}{*}{Distribution} &BBB &AUROC &XGBoost &0.907 ± 0.002 & &0.903 ± 0.005 &2 \\
&PPBR &MAE &XGBoost &8.252 ± 0.190 & &\underline{\textbf{8.073 ± 0.335}} &1 \\
&VDss &Spearman &XGBoost &0.627 ± 0.009 & &\underline{\textbf{0.654 ± 0.031}} &1 \\
\cmidrule{1-8}
\multirow{6}{*}{Metabolism} &CYP2D6 inhibition &AUPRC &XGBoost &0.717 ± 0.001 & &0.682 ± 0.008 &2 \\
&CYP3A4 inhibition &AUPRC &XGBoost &0.872 ± 0.005 & &0.867 ± 0.003 &2 \\
&CYP2C9 inhibition &AUPRC &XGBoost &0.769 ± 0.000 & &\underline{\textbf{0.801 ± 0.003}} &1 \\
&CYP2D6 substrate &AUPRC &RDKit2D + MLP (DeepPurpose) &0.677 ± 0.047 & &\underline{\textbf{0.699 ± 0.018}} &1 \\
&CYP3A4 substrate &AUROC &XGBoost &0.677 ± 0.007 & &0.670 ± 0.018 &2 \\
&CYP2C9 substrate &AUPRC &SimGCN &0.433 ± 0.017 & &\underline{\textbf{0.446 ± 0.062}} &1 \\
\cmidrule{1-8}
\multirow{3}{*}{Excretion} &Half life &Spearman &SimGCN &0.392 ± 0.065 & &\underline{\textbf{0.549 ± 0.024}} &1 \\
&Clearance microsome &Spearman &SimGCN &0.597 ± 0.025 & &\underline{\textbf{0.607 ± 0.027}} &1 \\
&Clearance hepatocyte &Spearman &ContextPred &0.439 ± 0.026 & &0.381 ± 0.038 &6 \\
\cmidrule{1-8}
\multirow{4}{*}{Toxicity} &hERG &AUROC &SimGCN &0.874 ± 0.014 & &0.823 ± 0.009 &4 \\
&Ames &AUROC &XGBoost &0.859 ± 0.000 & &0.813 ± 0.005 &8 \\
&DILI &AUROC &XGBoost &0.925 ± 0.012 & &0.883 ± 0.021 &5 \\
&LD50 &MAE &MACCS keys + autoML &0.588 ± 0.005 & &\underline{\textbf{0.577 ± 0.019}} &1 \\
\bottomrule
\end{tabular}
\end{table}

We hypothesize that learning atom environments will enforce the model to aggregate the local chemical groups that will be used for prediction. Learning an embedding of  atom environments and how to aggregate them into a molecular embedding can help to solve some problems of classical fingerprints such as sparsity and clashes when using bit vectors. Interestingly, this self-supervision approach is not limited to transformers since it can easily be used to pretrain GNN. It is still to be determined the effect of diversity and amount of data used during the self-supervision since we only used drug-like molecules included in GuacaMol training data. Nonetheless we expect that larger and more diverse datasets can only improve current performance of the model. Overall we consider this work as an initial step towards more powerful foundation models for chemical property prediction.

\bibliographystyle{unsrturl}
\bibliography{ref}

\newpage

\appendix
\section{Extended results}
\label{ext_results}

\begin{table}[!htp] 
    \caption{Results of fine-tuning MolE without any pretraining or with only supervised pretraining on the TDC ADMET tasks. Underlined values show cases where the model outperforms results on the TDC leaderboard (as of September 2022).} 
    \label{tabS1}
    \centering
    \scriptsize
    \begin{tabular}{lrcccc}
        \toprule
        &Dataset &Metric &\makecell{Best model in\\TDC Leaderboard} &\makecell{MolE\\No pretraining} &\makecell{MolE only\\Supervised pretraining} \\
        \midrule
\multirow{6}{*}{Absorption} &Caco2 &MAE &0.291 ± 0.015 &0.400 ± 0.016 &0.392 ± 0.025 \\
&HIA &AUROC &0.988 ± 0.002 &0.969 ± 0.004 &\underline{\textbf{0.988 ± 0.002}} \\
&Pgp &AUROC &0.929 ± 0.010 &0.848 ± 0.021 &0.896 ± 0.010 \\
&Bioavailability &AUROC &0.748 ± 0.033 &0.656 ± 0.026 &0.566 ± 0.066 \\
&Lipophilicity &MAE &0.535 ± 0.012 &0.624 ± 0.025 &0.586 ± 0.020 \\
&Solubility &MAE &0.734 ± 0.006 &0.808 ± 0.031 &0.843 ± 0.013 \\
\cmidrule(r){1-6}
\multirow{3}{*}{Distribution} &BBB &AUROC &0.907 ± 0.002 &0.854 ± 0.010 &0.872 ± 0.021 \\
&PPBR &MAE &8.252 ± 0.190 &8.790 ± 0.525 &8.949 ± 0.396 \\
&VDss &Spearman &0.627 ± 0.009 &0.607 ± 0.039 &\underline{\textbf{0.642 ± 0.013}} \\
\cmidrule(r){1-6}
\multirow{6}{*}{Metabolism} &CYP2D6 inhibition &AUPRC &0.717 ± 0.001 &0.601 ± 0.017 &0.620 ± 0.003 \\
&CYP3A4 inhibition &AUPRC &0.872 ± 0.005 &0.802 ± 0.028 &0.818 ± 0.005 \\
&CYP2C9 inhibition &AUPRC &0.769 ± 0.000 &0.710 ± 0.009 &0.736 ± 0.007 \\
&CYP2D6 substrate &AUPRC &0.677 ± 0.047 &0.576 ± 0.061 &0.665 ± 0.024 \\
&CYP3A4 substrate &AUROC &0.677 ± 0.007 &0.556 ± 0.017 &0.606 ± 0.018 \\
&CYP2C9 substrate &AUPRC &0.433 ± 0.017 &0.374 ± 0.047 &0.351 ± 0.013 \\
\cmidrule(r){1-6}
\multirow{3}{*}{Excretion} &Half life &Spearman &0.392 ± 0.065 &0.386 ± 0.043 &\underline{\textbf{0.477 ± 0.027}} \\
&Clearance microsome &Spearman &0.597 ± 0.025 &0.443 ± 0.019 &0.508 ± 0.022 \\
&Clearance hepatocyte &Spearman &0.439 ± 0.026 &0.336 ± 0.014 &0.413 ± 0.015 \\
\cmidrule(r){1-6}
\multirow{4}{*}{Toxicity} &hERG &AUROC &0.874 ± 0.014 &0.704 ± 0.026 &0.816 ± 0.012 \\
&Ames &AUROC &0.859 ± 0.000 &0.782 ± 0.009 &0.792 ± 0.005 \\
&DILI &AUROC &0.925 ± 0.012 &0.897 ± 0.006 &0.900 ± 0.008 \\
&LD50 &MAE &0.588 ± 0.005 &0.691 ± 0.024 &0.649 ± 0.017 \\
\bottomrule
\end{tabular}
\end{table}

\begin{table}[!htp] 
    \caption{Results of fine-tuning pretrained MolE on the ADMET group of the TDC benchmark. Underlined values show cases where the model outperforms results on the TDC leaderboard (as of September 2022).} 
    \label{tabS2}
    \centering
    \scriptsize
    \begin{tabular}{lrccccc}
        \toprule
        &Dataset &Metric &\makecell{MolE\\(AtomEnvs)} &\makecell{MolE\\(FunctionalEnvs)} &\makecell{MolE (AtomEnvs)\\+ Supervised} &\makecell{MolE (FunctionalEnvs)\\+ Supervised} \\
        \midrule
        \multirow{6}{*}{Absorption} &Caco2 &MAE &0.471 ± 0.049 &0.355 ± 0.025 &0.310 ± 0.010 &0.366 ± 0.063 \\
&HIA &AUROC &0.949 ± 0.004 &0.951 ± 0.016 &0.963 ± 0.019 &0.974 ± 0.011 \\
&Pgp &AUROC &0.871 ± 0.021 &0.873 ± 0.023 &0.915 ± 0.005 &0.902 ± 0.014 \\
&Bioavailability &AUROC &0.683 ± 0.011 &0.638 ± 0.027 &0.654 ± 0.028 &0.705 ± 0.029 \\
&Lipophilicity &MAE &\underline{\textbf{0.464 ± 0.008}} &\underline{\textbf{0.460 ± 0.007}} &\underline{\textbf{0.469 ± 0.009}} &\underline{\textbf{0.467 ± 0.006}} \\
&Solubility &MAE &0.810 ± 0.020 &0.799 ± 0.007 &0.792 ± 0.005 &0.771 ± 0.009 \\
\cmidrule(r){1-7}
\multirow{3}{*}{Distribution} &BBB &AUROC &0.895 ± 0.006 &0.901 ± 0.008 &0.903 ± 0.005 &0.903 ± 0.007 \\
&PPBR &MAE &\underline{\textbf{8.191 ± 0.189}} &8.570 ± 0.237 &\underline{\textbf{8.073 ± 0.335}} &8.379 ± 0.236 \\
&VDss &Spearman &0.596 ± 0.020 &0.622 ± 0.021 &\underline{\textbf{0.654 ± 0.031}} &\underline{\textbf{0.648 ± 0.024}} \\
\cmidrule(r){1-7}
\multirow{6}{*}{Metabolism} &CYP2D6 inhibition &AUPRC &0.665 ± 0.025 &0.678 ± 0.010 &0.682 ± 0.008 &0.693 ± 0.007 \\
&CYP3A4 inhibition &AUPRC &0.865 ± 0.005 &0.857 ± 0.006 &0.867 ± 0.003 &0.862 ± 0.003 \\
&CYP2C9 inhibition &AUPRC &\underline{\textbf{0.773 ± 0.006}} &0.759 ± 0.015 &\underline{\textbf{0.801 ± 0.003}} &\underline{\textbf{0.797 ± 0.004}} \\
&CYP2D6 substrate &AUPRC &\underline{\textbf{0.706 ± 0.023}} &\underline{\textbf{0.715 ± 0.011}} &\underline{\textbf{0.699 ± 0.018}} &\underline{\textbf{0.728 ± 0.025}} \\
&CYP3A4 substrate &AUROC &0.633 ± 0.012 &0.612 ± 0.014 &0.670 ± 0.018 &0.669 ± 0.012 \\
&CYP2C9 substrate &AUPRC &0.429 ± 0.092 &0.411 ± 0.027 &\underline{\textbf{0.446 ± 0.062}} &\underline{\textbf{0.452 ± 0.024}} \\
\cmidrule(r){1-7}
\multirow{3}{*}{Excretion} &Half life &Spearman &\underline{\textbf{0.518 ± 0.045}} &\underline{\textbf{0.579 ± 0.035}} &\underline{\textbf{0.549 ± 0.024}} &\underline{\textbf{0.537 ± 0.032}} \\
&Clearance microsome &Spearman &0.531 ± 0.024 &0.567 ± 0.016 &\underline{\textbf{0.607 ± 0.027}} &\underline{\textbf{0.598 ± 0.027}} \\
&Clearance hepatocyte &Spearman &0.367 ± 0.039 &0.373 ± 0.027 &0.381 ± 0.038 &0.381 ± 0.016 \\
\cmidrule(r){1-7}
\multirow{4}{*}{Toxicity} &hERG &AUROC &0.844 ± 0.021 &0.871 ± 0.007 &0.823 ± 0.009 &0.829 ± 0.004 \\
&Ames &AUROC &0.832 ± 0.019 &0.831 ± 0.017 &0.813 ± 0.005 &0.808 ± 0.009 \\
&DILI &AUROC &0.883 ± 0.021 &0.890 ± 0.016 &0.883 ± 0.021 &0.906 ± 0.016 \\
&LD50 &MAE &\underline{\textbf{0.582 ± 0.010}} &0.597 ± 0.027 &\underline{\textbf{0.577 ± 0.019}} &0.651 ± 0.032 \\
\bottomrule
\end{tabular}
\end{table}

\begin{table}[!htp] 
    \caption{Results of fine-tuning pretrained MolE using logP prediction as auxiliary loss during self-supervised pretraining. Underlined values show cases where the model outperforms results on the TDC leaderboard (as of September 2022).} 
    \label{tabS3}
    \centering
    \scriptsize
    \begin{tabular}{lrccccc}
        \toprule
        &Dataset &Metric &\makecell{MolE (AtomEnvs \\+ logP)} &\makecell{MolE (FunctionalEnvs\\+ logP)} &\makecell{MolE (AtomEnvs\\+ logP)+ Supervised} &\makecell{MolE (FunctionalEnvs\\+ logP) + Supervised} \\
        
    \midrule
\multirow{6}{*}{Absorption} &Caco2 &MAE &0.344 ± 0.013 &0.389 ± 0.049 &0.355 ± 0.049 &0.325 ± 0.027 \\
&HIA &AUROC &0.959 ± 0.005 &0.969 ± 0.013 &0.959 ± 0.006 &0.983 ± 0.007 \\
&Pgp &AUROC &0.880 ± 0.018 &0.867 ± 0.018 &0.919 ± 0.014 &0.921 ± 0.013 \\
&Bioavailability &AUROC &0.633 ± 0.014 &0.622 ± 0.038 &0.660 ± 0.034 &0.698 ± 0.046 \\
&Lipophilicity &MAE &\underline{\textbf{0.454 ± 0.006}} &\underline{\textbf{0.448 ± 0.006}} &\underline{\textbf{0.451 ± 0.014}} &\underline{\textbf{0.448 ± 0.009}} \\
&Solubility &MAE &0.775 ± 0.010 &0.776 ± 0.018 &0.758 ± 0.005 &0.770 ± 0.013 \\
\cmidrule(r){1-7}
\multirow{3}{*}{Distribution} &BBB &AUROC &0.874 ± 0.017 &0.880 ± 0.013 &0.889 ± 0.011 &\underline{\textbf{0.914 ± 0.012}} \\
&PPBR &MAE &\underline{\textbf{7.993 ± 0.124}} &\underline{\textbf{7.926 ± 0.255}} &8.605 ± 0.233 &\underline{\textbf{8.128 ± 0.209}} \\
&VDss &Spearman &\underline{\textbf{0.643 ± 0.020}} &0.612 ± 0.024 &\underline{\textbf{0.658 ± 0.014}} &\underline{\textbf{0.641 ± 0.019}} \\
\cmidrule(r){1-7}
\multirow{6}{*}{Metabolism} &CYP2D6 inhibition &AUPRC &0.666 ± 0.011 &0.655 ± 0.029 &0.700 ± 0.003 &0.697 ± 0.002 \\
&CYP3A4 inhibition &AUPRC &0.853 ± 0.007 &0.847 ± 0.004 &\underline{\textbf{0.876 ± 0.001}} &0.867 ± 0.010 \\
&CYP2C9 inhibition &AUPRC &0.758 ± 0.019 &0.765 ± 0.022 &\underline{\textbf{0.790 ± 0.004}} &\underline{\textbf{0.791 ± 0.004}} \\
&CYP2D6 substrate &AUPRC &0.645 ± 0.017 &\underline{\textbf{0.680 ± 0.019}} &\underline{\textbf{0.729 ± 0.015}} &0.648 ± 0.024 \\
&CYP3A4 substrate &AUROC &0.613 ± 0.019 &0.639 ± 0.016 &0.628 ± 0.026 &0.643 ± 0.013 \\
&CYP2C9 substrate &AUPRC &0.374 ± 0.069 &\underline{\textbf{0.442 ± 0.023}} &0.424 ± 0.067 &0.420 ± 0.013 \\
\cmidrule(r){1-7}
\multirow{3}{*}{Excretion} &Half life &Spearman &\underline{\textbf{0.492 ± 0.050}} &\underline{\textbf{0.543 ± 0.033}} &\underline{\textbf{0.533 ± 0.042}} &\underline{\textbf{0.529 ± 0.046}} \\
&Clearance microsome &Spearman &0.571 ± 0.024 &0.574 ± 0.034 &0.594 ± 0.012 &\underline{\textbf{0.615 ± 0.017}} \\
&Clearance hepatocyte &Spearman &0.376 ± 0.041 &0.374 ± 0.043 &0.417 ± 0.027 &0.407 ± 0.020 \\
\cmidrule(r){1-7}
\multirow{4}{*}{Toxicity} &hERG &AUROC &0.843 ± 0.022 &0.817 ± 0.012 &0.842 ± 0.010 &0.818 ± 0.024 \\
&Ames &AUROC &0.827 ± 0.007 &0.833 ± 0.008 &0.790 ± 0.017 &0.821 ± 0.008 \\
&DILI &AUROC &0.898 ± 0.007 &0.874 ± 0.088 &0.868 ± 0.013 &0.880 ± 0.018 \\
&LD50 &MAE &0.597 ± 0.018 &0.605 ± 0.035 &0.603 ± 0.013 &0.608 ± 0.005 \\
\bottomrule
\end{tabular}
\end{table}

\begin{table}[!htp] 
    \caption{Results of fine-tuning pretrained MolE using fingerprint prediction as auxiliary loss during self-supervised pretraining. Underlined values show cases where the model outperforms results on the TDC leaderboard (as of September 2022).} 
    \label{tabS4}
    \centering
    \scriptsize
    \begin{tabular}{lrccccc}
        \toprule
        &Dataset &Metric &\makecell{MolE (AtomEnvs \\+ FP)} &\makecell{MolE (FunctionalEnvs\\+ FP)} &\makecell{MolE (AtomEnvs\\+ FP)+ Supervised} &\makecell{MolE (FunctionalEnvs\\+ FP) + Supervised} \\
        
    \midrule
\multirow{6}{*}{Absorption} &Caco2 &MAE &0.371 ± 0.032 &0.341 ± 0.021 &0.331 ± 0.039 &0.372 ± 0.014 \\
&HIA &AUROC &0.939 ± 0.009 &0.956 ± 0.013 &0.972 ± 0.007 &0.945 ± 0.007 \\
&Pgp &AUROC &0.891 ± 0.016 &0.905 ± 0.011 &0.902 ± 0.004 &0.906 ± 0.009 \\
&Bioavailability &AUROC &0.674 ± 0.04 &0.662 ± 0.026 &0.646 ± 0.059 &0.678 ± 0.046 \\
&Lipophilicity &MAE &\underline{\textbf{0.473 ± 0.007}} &\underline{\textbf{0.457 ± 0.009}} &\underline{\textbf{0.48 ± 0.011}} &\underline{\textbf{0.483 ± 0.017}} \\
&Solubility &MAE &0.825 ± 0.017 &0.79 ± 0.012 &0.803 ± 0.012 &0.789 ± 0.01 \\
\cmidrule(r){1-7}
\multirow{3}{*}{Distribution} &BBB &AUROC &0.891 ± 0.032 &0.904 ± 0.021 &0.903 ± 0.008 &0.895 ± 0.038 \\
&PPBR &MAE &8.507 ± 0.289 &8.392 ± 0.181 &\underline{\textbf{8.224 ± 0.183}} &\underline{\textbf{8.105 ± 0.175}} \\
&VDss &Spearman &0.605 ± 0.043 &0.611 ± 0.024 &\underline{\textbf{0.645 ± 0.033}} &\underline{\textbf{0.662 ± 0.009}} \\
\cmidrule(r){1-7}
\multirow{6}{*}{Metabolism} &CYP2D6 inhibition &AUPRC &0.677 ± 0.014 &0.676 ± 0.018 &0.668 ± 0.013 &0.687 ± 0.007 \\
&CYP3A4 inhibition &AUPRC &0.861 ± 0.005 &0.87 ± 0.006 &\underline{\textbf{0.874 ± 0.004}} &0.867 ± 0.006 \\
&CYP2C9 inhibition &AUPRC &0.753 ± 0.01 &\underline{\textbf{0.781 ± 0.003}} &\underline{\textbf{0.784 ± 0.003}} &\underline{\textbf{0.8 ± 0.003}} \\
&CYP2D6 substrate &AUPRC &\underline{\textbf{0.712 ± 0.018}} &\underline{\textbf{0.698 ± 0.031}} &0.669 ± 0.026 &\underline{\textbf{0.713 ± 0.042}} \\
&CYP3A4 substrate &AUROC &0.639 ± 0.021 &0.648 ± 0.01 &0.652 ± 0.013 &0.664 ± 0.007 \\
&CYP2C9 substrate &AUPRC &0.41 ± 0.027 &0.415 ± 0.053 &\underline{\textbf{0.459 ± 0.015}} &\underline{\textbf{0.443 ± 0.016}} \\
\cmidrule(r){1-7}
\multirow{3}{*}{Excretion} &Half life &Spearman &\underline{\textbf{0.523 ± 0.052}} &\underline{\textbf{0.517 ± 0.049}} &\underline{\textbf{0.515 ± 0.065}} &\underline{\textbf{0.58 ± 0.03}} \\
&Clearance microsome &Spearman &0.537 ± 0.024 &0.532 ± 0.016 &0.55 ± 0.023 &0.572 ± 0.016 \\
&Clearance hepatocyte &Spearman &0.403 ± 0.038 &0.4 ± 0.018 &0.384 ± 0.04 &0.399 ± 0.042 \\
\cmidrule(r){1-7}
\multirow{4}{*}{Toxicity} &hERG &AUROC &0.824 ± 0.007 &0.846 ± 0.022 &0.827 ± 0.023 &0.824 ± 0.011 \\
&Ames &AUROC &0.831 ± 0.005 &0.831 ± 0.009 &0.786 ± 0.012 &0.808 ± 0.004 \\
&DILI &AUROC &0.843 ± 0.093 &0.844 ± 0.039 &0.878 ± 0.012 &0.889 ± 0.013 \\
&LD50 &MAE &0.621 ± 0.035 &0.593 ± 0.009 &0.599 ± 0.025 &0.619 ± 0.007 \\
\bottomrule
\end{tabular}
\end{table}

\end{document}